\documentclass[twocolumn]{aastex701}

\usepackage{soul}

\newcommand{\bxc}{\textsc{BxC}}

\newcommand{\ffig}{Figure~}

\newcommand\bb[1]{\mbox{\boldmath{$#1$}}}
\newcommand\grad{\bb{\nabla}}
\newcommand\bcdot{\,\bb{\cdot}\,}

\newcommand\btimes{\,\bb{\times}\,}

\begin{document}

\title{High-energy Particle Transport in Three-dimensional Anisotropic Turbulent Magnetic Fields}

\author[0000-0002-3392-8329]{Daniela Maci}
\affiliation{Centre for mathematical Plasma Astrophysics, Department of Mathematics, KU Leuven, Celestijnenlaan 200B, B-3001 Leuven, Belgium}
\email[show]{daniela.maci@kuleuven.be}  

\author[0000-0003-3544-2733]{Rony Keppens}
\affiliation{Centre for mathematical Plasma Astrophysics, Department of Mathematics, KU Leuven, Celestijnenlaan 200B, B-3001 Leuven, Belgium}
\email[]{rony.keppens@kuleuven.be}  

\author[0000-0002-7526-8154]{Fabio Bacchini}
\affiliation{Centre for mathematical Plasma Astrophysics, Department of Mathematics, KU Leuven, Celestijnenlaan 200B, B-3001 Leuven, Belgium}
\affiliation{Royal Belgian Institute for Space Aeronomy, Solar-Terrestrial Centre of Excellence, Ringlaan 3, 1180 Uccle, Belgium}
\email[]{fabio.bacchini@kuleuven.be}

\begin{abstract}
The understanding and modeling of high-energy particles transport in turbulent magnetic fields is an important open question in space- and astrophysics. The multiscale, nonlinear nature of turbulence, and the high variability of turbulence properties across different environments, make it particularly challenging to reach a full understanding of the interactions between particles and turbulent fluctuations. Using synthetic, realistically looking turbulent magnetic field realizations generated by the \bxc{} toolkit, we investigate how the scattering of particles is affected by anisotropic fluctuations in strongly turbulent fields. We find evidence that, in the absence of a uniform background or guide magnetic field, the scattering process is not governed by the turbulence correlation length. We then further verify this hypothesis by studying particle transport in the presence of a guide field. We find evidence of a different scattering mechanism than the usual pitch-angle diffusion used to describe scattering in strong-guide-field settings.  
\end{abstract}


\section{Introduction} 
The modeling of high-energy-particle transport is a long-standing problem of broad interest in space and astrophysics. It has direct implications for the acceleration and propagation of particles in the heliosphere, as well as in the interstellar and intergalactic media. Despite the widespread attention and effort devoted to this topic (see e.g.\ \citealt{cr_review} for a review), a complete understanding of the scattering process is still lacking. This process is strictly related to the turbulent magnetic fluctuations that permeate astrophysical environments and, in principle, requires a full description of the interactions between particles and fluctuations, most commonly described in the framework of magnetohydrodynamics (MHD) \citep{interactions}. Achieving such an understanding has proven challenging because of the multiscale and nonlinear nature of turbulence, as well as the large variability of its properties across different environments, most of which are still the subject of debate in the scientific community (see e.g.\ \citealt{MHD_review} for a review). 

Considerable efforts have been devoted to building theoretical models describing particle transport and scattering as functions of turbulence properties. Quasilinear theory (QLT) has been rather successful in predicting parallel diffusion coefficients \citep{qlt1, qlt2, qlt3, qlt4}, but the perturbative approach intrinsic to this theory is based on the assumption of ``unperturbed orbits'', meaning that its validity is limited to environments with strong background fields that are combined with small-amplitude fluctuations. Additionally, a well-known limitation of QLT is the \textit{90-degree} problem: for vanishing pitch angle $\mu$ (i.e.\ when the particle velocity is perpendicular to the background magnetic field), QLT predicts a vanishing pitch-angle diffusion coefficient, hence particles cannot scatter efficiently from $\mu < 0$ to $\mu > 0$. However, it has been established through simulations (see \citet{tp_review2020} for a review) that this is an artifact of the theory, probably due to the simplifying assumptions behind QLT. While some limitations of QLT have been addressed through nonlinear extensions \citep{bam1997, nlgc2003, weaklynl2004, shalchi2005, tautz2008, shalchi2009}, no comprehensive framework exists that self-consistently captures high-energy particle transport and scattering in the presence of generic turbulent fluctuations across all relevant spatial scales. 

In simulations, the propagation of high-energy particles is often treated within the magnetostatic approximation, i.e.\ in a static turbulent background. While most studies agree in treating particles as test particles (i.e.\ without considering the feedback of particles onto the fields), the generation of the background turbulent fields is in itself approached from different angles. On the one hand, many studies use synthetic fields (i.e.\ constructed ad hoc based on turbulence statistical properties) and have achieved great success in the endeavor of exploring the effects of turbulent magnetic fluctuations on particle scattering \citep{Casse2001, bibi2014, pucci2016, Snodin2016, Snodin2017, Dundovic2020, Reichherzer2020, tp_review2020, Els2024, lubke24, Pezzi2024}. Synthetic turbulent fields focus on the reproduction of turbulent magnetic fluctuations without actually handling the nonlinear MHD evolution. At the same time, employing synthetic models is justified by the minimal amount of computational resources they require. Specifically, using synthetic fields allows for exploring the effects of a wide range of turbulence-related properties, such as covering a significantly extended inertial range \citep{pucci2016}, or variations in the amplitude of magnetic fluctuations compared to ordered background fields \citep{Snodin2016, Dundovic2020, Reichherzer2020, Els2024}. 

On the other hand, a series of studies have used turbulent magnetic fields from direct numerical simulations (DNSs) as a background for the propagation of particles \citep{Chandran2000, mhd_th2, cohet, mhd_sim1, mhd_sim2}. DNSs have the undisputed advantage of generating a realistic environment that models the full properties of MHD-governed physical systems, including electric and velocity fields, currents, etc., as well as controlling plasma parameters such as the plasma-$\beta$ and Mach number, which have been shown to affect the propagation of high-energy particles \citep{plasma_beta1, plasma_beta2, cohet}. However, despite recent improvements in computational resources and techniques, DNSs are still dramatically more expensive than synthetic models, to the point that parametric studies surveying a wide range of turbulence properties would be prohibitive. For this reason, studies involving synthetic turbulence are still widely used and new models are being developed in order to include more realistic features. For instance, particular attention has recently been given to the effects of intermittent fields compared to statistically Gaussian fluctuations \citep{bibi2014, pucci2016, Snodin2017, Lemoine2023}. 

It is now established that astrophysical turbulence is mostly anisotropic, but it has also been shown that not all astrophysical environments exhibit the same features. Anisotropy is especially important in particle transport, as it determines scattering parallel and perpendicular to a background magnetic field, but a comprehensive theory of transport in anisotropic turbulence is still lacking. A complicating factor is that anisotropy is not uniquely defined, but it can manifest in different forms. A turbulent field can show spectral anisotropy, i.e.\ with power unevenly distributed across different scales, and/or variance anisotropy, i.e.\ with power unevenly distributed among the different field components. Additionally, anisotropy might be present at large (global) scales, usually due to the presence of a strong guide field, or locally.  Another challenging aspect, is that the anisotropy of astrophysical turbulence is still uncertain and appears to vary substantially from one environment to another. In the heliosphere, solar-wind turbulence is well-established to be strongly anisotropic with respect to the local mean magnetic field, but the specific nature of this anisotropy is still under debate. The solar corona is also widely believed to host anisotropic Alfvénic turbulence, especially in open-field regions \citep{intro2}. For Jupiter, there is specific evidence that turbulence in the middle magnetosphere can lie in the regime of weak MHD turbulence \citep{jupiter}. The interstellar medium is less uniform and its degree of anisotropy strongly depends on phase, scale, and magnetization \citep{ism2, ism1, ism3}.

Several studies including anisotropic fluctuations have been performed on synthetic fields in which anisotropy is obtained through a composite model of a 1D-slab and a 2D perpendicular components \citep{Dundovic2020}. Although this composite model is numerically practical and it is observationally motivated \citep{Matt1990}, it is not representative of anisotropic turbulence in general. Another widely investigated model is the Goldreich--Sridhar \citep{GS95} anisotropic turbulence. The hypothesis of a Goldreich--Sridhar-type phenomenology is supported by several authors \citep{GS95, intro1, intro2}, while others \citep{Telloni19} found no evidence of critical-balance signatures in observational data of field-aligned solar-wind turbulence. At the same time, other studies \citep{Chandran2000, mhd_th2} have concluded that this specific model of turbulence leads to inefficient scattering, which therefore does not account for the observed values of confinement and acceleration of cosmic rays. In this context, studies conducted with MHD simulations rely on a self-consistently built anisotropy that is often more representative of various astrophysical scenarios (e.g.\ \citealt{cohet, mhd_sim2}). What is still sorely missing is a detailed study of the effects of different anisotropy prescriptions that have not been yet included in other synthetic models and it is important for a full understanding of particles transport in different environments. 

In this Letter, we perform test-particle simulations in synthetic turbulent magnetic fields in which anisotropy is built self-consistently in a fully 3D setting. In addition, the fields are intermittent and include coherent structures. To the best of our knowledge, this is the first time that all these fundamental aspects are fully incorporated into a synthetic model. By construction, in our approach the anisotropy level is fully controllable, which allows for anisotropic models other than the Goldreich--Sridhar one. We focus our study on highly turbulent settings, with low levels of anisotropy. We show that the mechanisms that govern particle scattering in highly turbulent environments are not the same as the ones that have been shown to govern scattering in strong-guide-field environments, suggesting that as the turbulence level increases, field-line curvature assumes a central role in the scattering process. Furthermore, we investigate to what extent such scattering mechanisms are significant, and we provide strong evidence that they contribute to particle scattering even at finite $\delta B/ B$ values. 

\section{Synthetic-Turbulence Model} We generate synthetic turbulent magnetic fields using the open-source, Python-based \bxc{} toolkit \citep{durrive, bxc, bxc_mpi}\footnote{https://bxc.academy}. Contrary to most synthetic models that use superposition of Fourier modes or wavelet-based methods \citep{wavelet2, wavelet3, fourier1, fourier2, Subedi_2014, wavelet1,  Lubke1}, \bxc{} produces grid-based turbulent magnetic fields using a combination of nonlinear geometric transformations on a white-noise vector field that enter the fundamental Biot--Savart law. The model's input parameters allow for direct control over the power-spectrum features (injection/dissipation scale, spectral exponent, etc.). Intermittency is introduced through a nonlinear transformation of a Gaussian field, which at the same time imposes typical geometric structures of turbulence (e.g.\ the appearance of weak to strong, geometrically structured current sheets). Previous validations against direct numerical simulations include the role of \bxc{}'s input parameters on higher-order structure functions \citep{durrive}. \bxc{} can reproduce statistical anisotropic fluctuations in 3D through direction-dependent parameters \citep{bxc}. The introduction of anisotropy in the fields is intrinsically geometric, and it is independent of the presence  of a global background field, which can be spatially varying or uniform, and is easily added in the model.

\begin{figure}
    \centering
    \includegraphics[width=\linewidth]{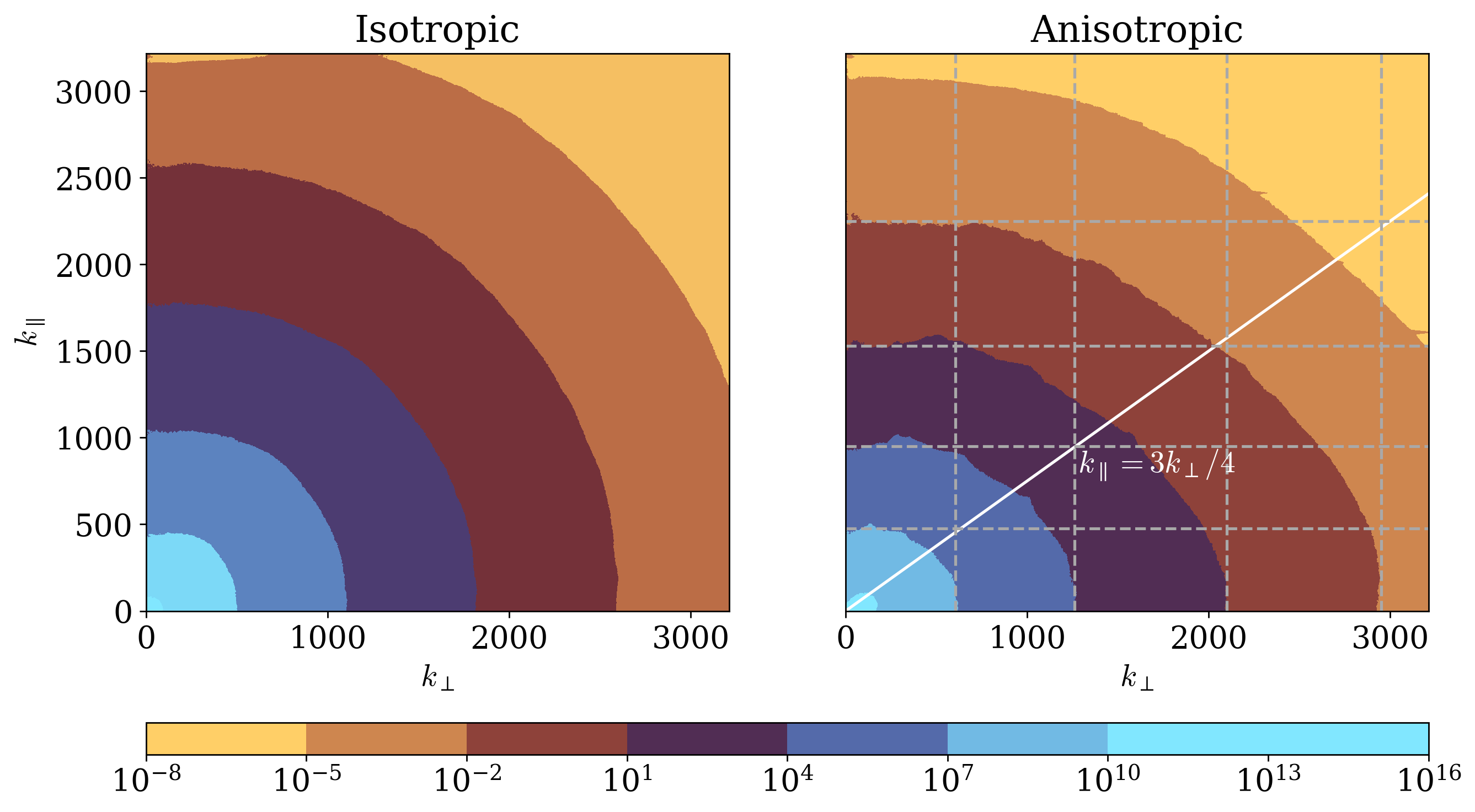}
    \caption{2D power spectrum of the turbulent isotropic (left) and anisotropic (right) field. Black lines are isocontours of the spectrum, plotted at the same levels on both fields.}
    \label{fig:ps}
\end{figure}

\begin{figure}
    \centering
    \includegraphics[width=\linewidth]{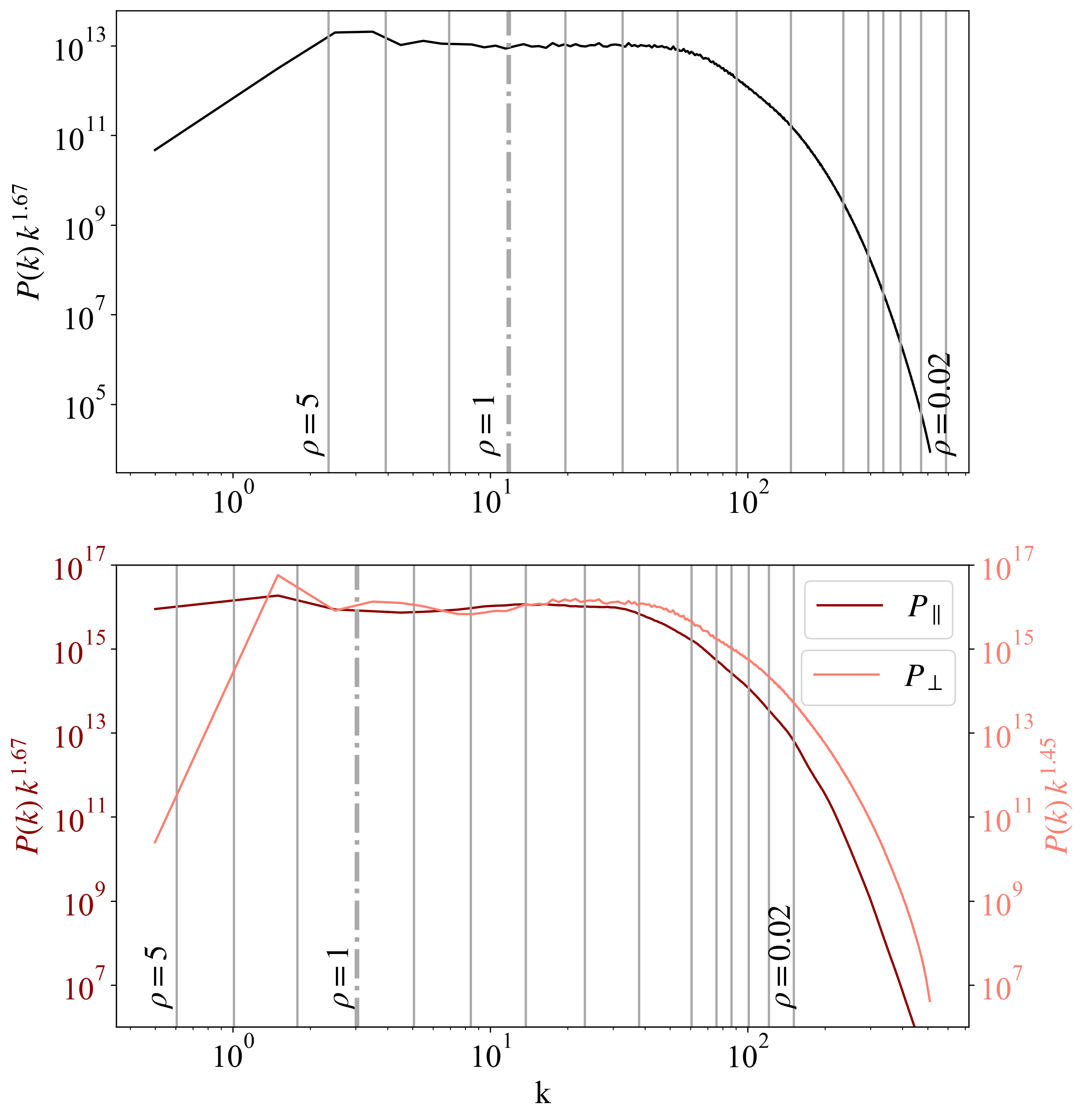}
    \caption{1D compensated power spectra for the isotropic (top panel) and anisotropic field (bottom panel). Vertical lines indicate the different values of gyroradius (i.e. different reduced rigidities) used to initialize the test particles.}
    \label{fig:ps1d}
\end{figure}

Here, we investigate the effect of both anisotropic fluctuations and background fields. The aim of this Letter is not to reproduce a specific astrophysical setting, but to show the capabilities of the model as well as to probe into turbulent regimes that are not usually investigated by means of synthetic models, in order to show the significance of understanding particle transport in different environments. For this reason, in our analysis we compare isotropic and weakly anisotropic fluctuations in highly turbulent settings. The isotropic case is chosen to follow a Kolmogorov-like scaling of the inertial range. \ffig\ref{fig:ps} shows the 2D power spectrum of the magnetic fluctuations (without background field) for the isotropic and anisotropic cases. The latter satisfies the linear relation $k_{\parallel} = 3k_{\perp}/4$, which indicates that the anisotropy is not scale-dependent. \ffig\ref{fig:ps1d} shows the integrated 1D spectra for the isotropic and anisotropic cases.  We generate turbulent fields in a box of unit length, on a grid of $N = 1024^3$ cells, with and without a uniform background field $\bb{B}_0 = B_0 \hat{\bb{z}}$. The field is, by construction, periodic in all three directions. When a background field is incorporated, the turbulent fluctuations are such that $\delta B/B_0 = 1$. Note that our study requires minimal computational resources compared to what a DNS would require. For example, \citet{cohet} estimate that, for one run, they use from ten hours to a few tens of hours of CPU computing time on a supercomputer. Our fields are generated in around 15 minutes on a regular desktop, and the particle propagation varies from 20 minutes to a few hours. 

\section{Test-particle simulation framework and set-up} The test-particle simulations are performed within a new framework combining \bxc{} and MPI-AMRVAC 3.0\footnote{https://amrvac.org} \citep{amrvac3, amrvac4}, an open-source, general-purpose partial differential equation solver equipped with a test-particle module that has been used in various studies of particle acceleration in solar eruptions \citep{Fabio2024,Mora2025,Hao2025}. Improving upon the procedure described in \citet{bxc_mpi}, we now use the Python package \texttt{Simesh}\footnote{https://github.com/Astery0502/simesh} to save the \bxc-generated field in a \verb+.dat+ file that matches the data structure of MPI-AMRVAC. This allows us to directly use the restart functionality of MPI-AMRVAC to simulate test particles with \bxc-generated fields and limit memory usage. 
The test-particle simulations are performed by solving the full relativistic equations of motion for charged particles using a standard Boris integrator \citep[e.g.,][]{ripperda2018}. As appropriate for highly energetic particles, here we use the magnetostatic approximation, i.e.\ we neglect the electric field ($\bb{E} = \bb{0}$) and the temporal evolution of the magnetic field ($\partial_t \bb{B} = \bb{0}$). 

For each field configuration, we consider 15 different values of normalized rigidities $\rho=r_g/\ell_C\in [2\times10^{-2}, 5]$, where $r_g$ is the initial particle gyroradius and $\ell_C$ the turbulence correlation length, which is computed numerically as the distance at which the autocorrelation function goes below the $1/e$ threshold. Specifically for the fields considered in this study, we have $\ell_{C, iso} = 0.085L$ and $\ell_{C, \parallel} = 0.3L$. The vertical lines in \ffig\ref{fig:ps1d} indicate the gyroradii used, and how they relate to the turbulence spectrum. For reference, the dashed line highlights the scale at which $\rho = 1$. For each run, we initialize $N_p=15{,}000$ particles with uniform velocity and random pitch angle. The particles velocity is determined from $r_g$, according to $v = |q|B_{rms}r_g/m$, using the root-mean-square of the total magnetic field and the mass-to-charge ratio equal to one.

\section{Results and Discussion}
We focus on how the scattering of particles is affected by the anisotropy of the fields. Specifically, we determine the diffusion coefficients ($D_{\parallel, \perp}$) numerically by taking the long-time limit of the running diffusion coefficients. The time it takes to reach the diffusive regime varies according to the velocity at which the particles are injected, which is determined on the basis of the particle gyroradii. For all cases considered, we made sure that the long-time limit of the running diffusion coefficient was stable for at least one decade in code-unit time. Our results are presented in terms of the effective mean free paths $\lambda_{\parallel, \perp} = 3 D_{\parallel, \perp} / v$, in order to isolate the scattering effects from the dynamical ones (i.e.\ the effect of the different particle velocities $v$). 

\begin{figure}
    \centering
    \includegraphics[width=\linewidth]{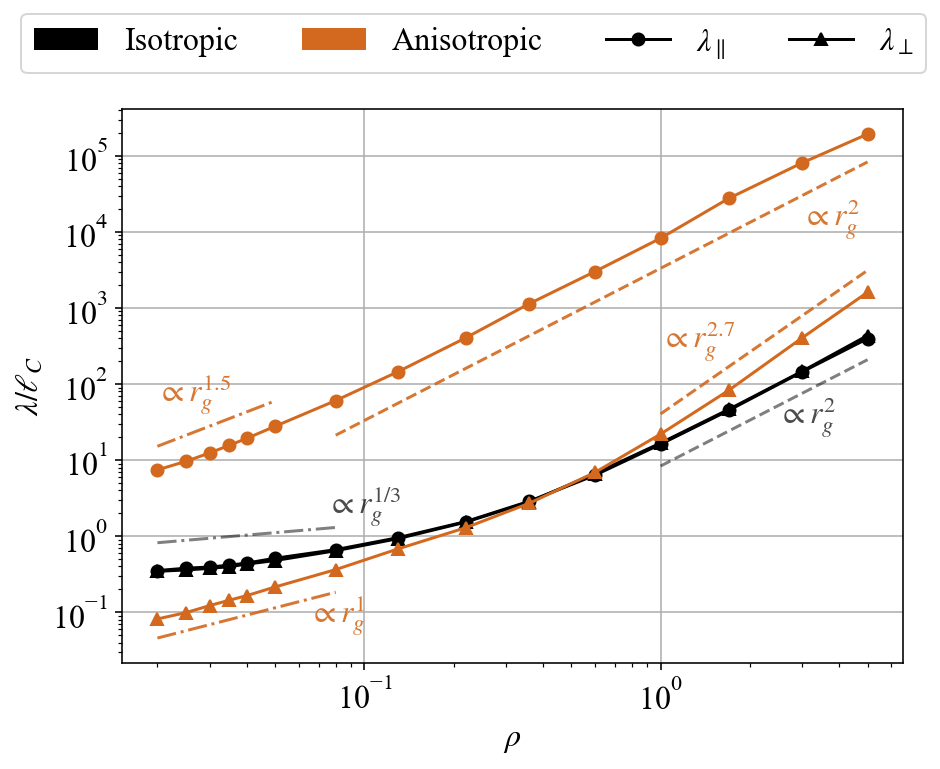}
    \caption{Parallel (dots) and perpendicular (triangles) mean free path as a function of normalized rigidity for isotropic (black) and anisotropic (orange) turbulent fields. In both isotropic and anisotropic fields adopted here, there is no background field.}
    \label{fig:nob0}
\end{figure}

First, we focus on the fully isotropic case. As this specific case has been explored in previous studies (see e.g.\ \citealt{Dundovic2020}), it allows us to benchmark the results obtained within the combined \bxc{}--AMRVAC framework against the expected scalings. In \ffig\ref{fig:nob0}, we show how $\lambda_{\parallel}$ and $\lambda_{\perp}$ scale as a function of $\rho$. As expected, in the fully isotropic case (black lines) there is no difference in the parallel and perpendicular directions (i.e.\ no preferential scattering direction). We also observe that the dependence of $\lambda$ on $\rho$ is consistent with what was found in previous studies, hence validating our framework: for $\rho \gtrsim 1$, $\lambda$ scales as $r_g^2$. This is considered to be an asymptotic scaling, expected when the particle gyroradius approaches and exceeds the characteristic scale of turbulence, and scattering due to turbulent fluctuations becomes less efficient \citep{plotnikov2011}. For $\rho \ll 1$, we also find the expected scaling of $\lambda \propto r_g^{1/3}$ that was recovered in previous studies (e.g.\ \citet{Dundovic2020}).

We now compare the scattering due to isotropic and anisotropic fluctuations without a background field. The results are shown in \ffig\ref{fig:nob0}, where $\lambda$ is plotted as a function of $\rho$. Significant differences can be noticed in $\lambda_{\parallel}$ (dots) and $\lambda_{\perp}$ (triangles) both with respect to each other and in comparison to the isotropic case. Let us first focus on the parallel scattering. We can observe that $\lambda_{\parallel}$ starts scaling as $r_g^{2}$ much earlier than expected ($\rho \lesssim 10^{-1}$). As mentioned above, $r_g^2$ is considered to be an asymptotic scaling expected when the particle gyroradius decorrelates from the turbulence scale responsible for the scattering. In the usual model of turbulence, this scale is taken as the turbulence correlation length: then, the mean free path asymptotically scales as $\lambda \propto r_g^2$ when the particle gyroradius is large enough that particles travel through uncorrelated regions of turbulence (at approximately $\rho \gtrsim 1$, as we find in the isotropic case). Our results contrast this picture: when anisotropy is introduced, we find that parallel scattering becomes inefficient at much lower values of $\rho$ than expected. Only for the lowest values of rigidities considered in this study we observe a slightly shallower scaling ($\lambda \propto r_g^{1.5}$), which is indicative of a transition region toward more efficient scattering at smaller scales. Our results indicate that, in our model, scattering is not regulated by the turbulence correlation length and that some other length scale, which is directly affected by the introduction of the anisotropy, is responsible instead. 

Before providing an explanation for what such a length scale could be, we analyze the results of perpendicular scattering. The behavior of $\lambda_{\perp}$ is less straightforward to interpret. We observe three different regimes: for small rigidities ($\rho \ll 1$), we find $\lambda_{\perp}\propto r_g^1$; for intermediate rigidities ($\rho < 1$), no clear power-law behavior is recognizable; and for large rigidities ($\rho \gtrsim 1$), we find $\lambda_{\perp} \propto r_g^{2.7}$. The intermediate rigidities clearly represent a transition region. It is unclear, however, whether the $r_g^{2.7}$ scaling is also part of a transition regime that will eventually relax to the asymptotic $r_g^2$ scaling, or whether the scaling will remain asymptotically the same. In any case, considering the high values of $\lambda$ it is associated to, as well as the steep slope it follows, it is a safe conclusion that this scattering process is not acting efficiently at these intermediate scales. The linear scaling at low rigidities bears resemblance to Bohm-like diffusion, which is often assumed in strong turbulence ($\delta B \gg B_0$) and used in shock acceleration models in the form $\lambda \propto  r_g$ \citep{shock1, shock2, shock3}. A linear scaling between $\lambda_{\perp}$ and $r_g$ has also been found in previous studies, both for synthetic turbulence \citep{bohm_shalchi}, and MHD-based \citep{cohet, mhd_sim2}. \citet{bohm_shalchi} specifically investigated the validity of the Bohm limit in highly turbulent synthetic fields, reaching values of $\delta B /B = 100$. However, they find $\lambda_{\perp} \propto r_g$, independently of the geometry of the field, whereas we find this result as a consequence of the introduction of anisotropy in the fluctuations. 

\begin{figure}
    \centering
    \includegraphics[width=\linewidth]{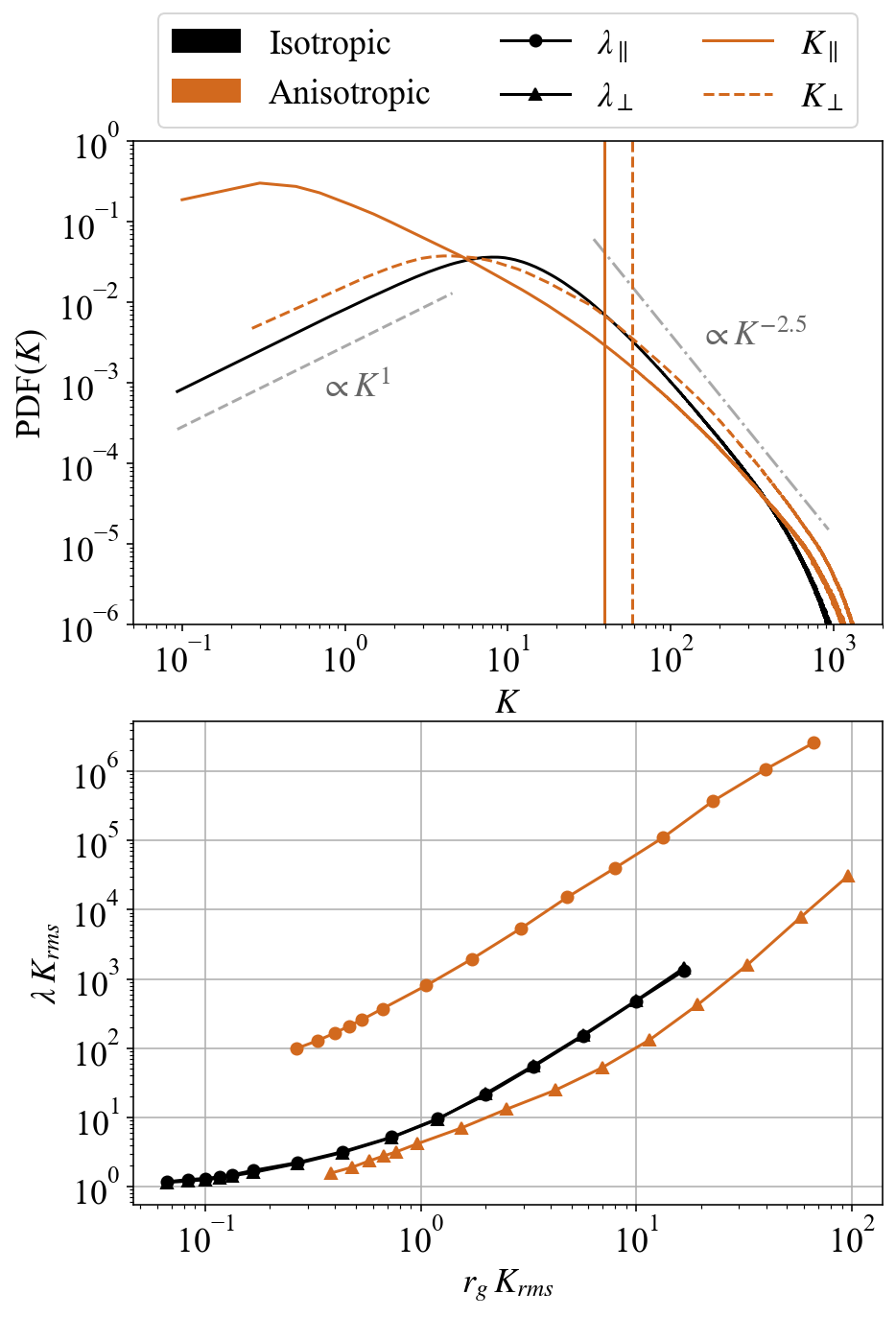}
    \caption{Top: Distribution function of field-line curvature $K_{\parallel}$ and inverse perpendicular reversal scale $K_{\perp}$ of the anisotropic field (orange lines). $K_{iso}$ (black line) corresponds to the field-line curvature of the isotropic field. Vertical lines indicate the root-mean-square of the vector fields. Bottom: parallel (circles) and perpendicular (triangles) mean free path as a function of gyroradius normalized by $K_{peak}^{-1}$ for isotropic (black) and anisotropic (orange) turbulent fields.}
    \label{fig:curv}
\end{figure}

A study featuring higher-resolution turbulent magnetic fields will be necessary to conclusively determine the nature of the scaling and have a complete picture of the efficient scattering in both parallel and perpendicular directions. However, this is out of the scope of this Letter and it is left for future work, as our main focus here is on the comparison between scattering by isotropic and anisotropic fluctuations. From the data available from this study, we can conclude that the anisotropy in the fluctuations significantly affects the regime of particle transport. In particular, scattering in the parallel direction becomes inefficient even for low rigidity values, whereas the scales of efficient scattering in the perpendicular direction are comparable to the isotropic case. This suggests that the length scale responsible for the scattering of particles is direction-dependent (i.e.\ it varies for the parallel and perpendicular directions) when anisotropy is introduced. In a recent study, \citet{kempski} have argued that, unlike the commonly studied strong-guide-field case, particle transport in strong magnetic turbulence may be regulated by small-scale intermittent field reversals and regions of strong field-line curvature. In this picture, transport differs qualitatively from the strong-guide-field limit because particles can be efficiently spatially scattered even while following magnetic-field lines, owing to the random walk of the folded field itself. Lower-energy particles are then more strongly confined because they can more closely follow these reversals, while they can also undergo additional scattering in regions where the field-line curvature becomes comparable to their gyroradius. Although a more detailed study would be required to conclusively determine the nature of the scattering processes in our model, here we propose an analysis that strongly suggests that our results are in agreement with what is conjectured by \citet{kempski}. In their analysis, \citet{kempski} make use of the field-line curvature $\bb{K}_{\parallel} \equiv \hat{\bb{b}} \bcdot \grad \hat{\bb{b}}$ and the inverse perpendicular reversal scale $\bb{K}_{\perp} = \hat{\bb{b}}\btimes (\hat{\bb{b}} \btimes \grad \log B)$, where $\hat{\bb{b}}$ is the unit vector along the local magnetic field. They show that in large-amplitude turbulence, the distributions of $|\bb{K}_{\parallel}|$ and $|\bb{K}_{\perp}|$ peak at different scales, contrary to what would happen in strong-guide-field turbulence. The top panel of \ffig\ref{fig:curv} shows the distribution of $K_{\parallel}$ and $K_{\perp}$ for the anisotropic case (orange lines) and the distribution of $K_{iso} = K_{\parallel, iso}$ for the isotropic case. The distribution of $K_{\perp, iso}$ is not shown because, as expected, it overlaps with that of $K_{\parallel, iso}$ almost exactly. In both the isotropic and anisotropic case, the curvature distribution has been computed for the purely turbulent fields (i.e.\ no background field included). The gray dashed lines indicate the scalings of the distribution function of the isotropic case. The scaling matches very well with those found by \citet{Yang19}, showing that our results are consistent with 3D MHD simulations. Similarly to \citet{Yang19}, we take the root-mean-square (RMS) of the curvature vector field as a reference scale, shown by the the vertical lines in \ffig\ref{fig:curv}. Note that the black line is not visible because it overlaps with the solid orange line. We then take the value of $K_{rms}^{-1}$ for each case (isotropic, anisotropic parallel, and anisotropic perpendicular) as a reference scale responsible for the scattering of particles. The bottom panel of \ffig\ref{fig:curv} shows the same results shown in \ffig\ref{fig:nob0}, but as a function of $r_g$ normalized here by $K_{rms}^{-1}$ instead of $\ell_C$. With this new normalization, the scaling regimes are more coherent with the standard physical interpretation, and more consistent with each other. In all cases (isotropic, anisotropic parallel, and anisotropic perpendicular), scattering starts being efficient efficient for $r_gK_{rms} \lesssim 10^{0}$, and $\lambda$ displays a clear power-law behavior. This is followed by a transition region whose extent varies from case to case, and eventually $\lambda$ follows the asymptotic scaling for $r_gK_{rms} > 1$, when the particle gyroradius has exceeded the characteristic curvature length scale. A higher-resolution study would be needed to confirm that the mean free path of the anisotropic case will also show efficient scattering in the same regime as the isotropic case. However, we believe that these results already provide strong evidence that, consistent with \citet{kempski, Yang19}, field-line curvature plays a much more decisive role than the turbulence correlation length in particle scattering under highly turbulent conditions. Note that we have also considered the possibility that the peak of the curvature sets the relevant reference scale, following what was suggested by \citet{kempski}. However, we find that the RMS curvature gives better results in the interpretation of scattering regimes. This may be due to the fact that the RMS value is more representative of infrequent, large values of curvature that would affect the particles scattering more than the ``most common'' curvature value set by the peak of the distribution. Such result would then be directly tied to the intermittent character of the turbulent fluctuations. In order to fully determine whether intermittency is at the root of these observations, a detailed study featuring a direct comparison between Gaussian and intermittent fluctuations is needed, and it is left for future work.

\begin{figure}
    \centering
    \includegraphics[width=\linewidth]{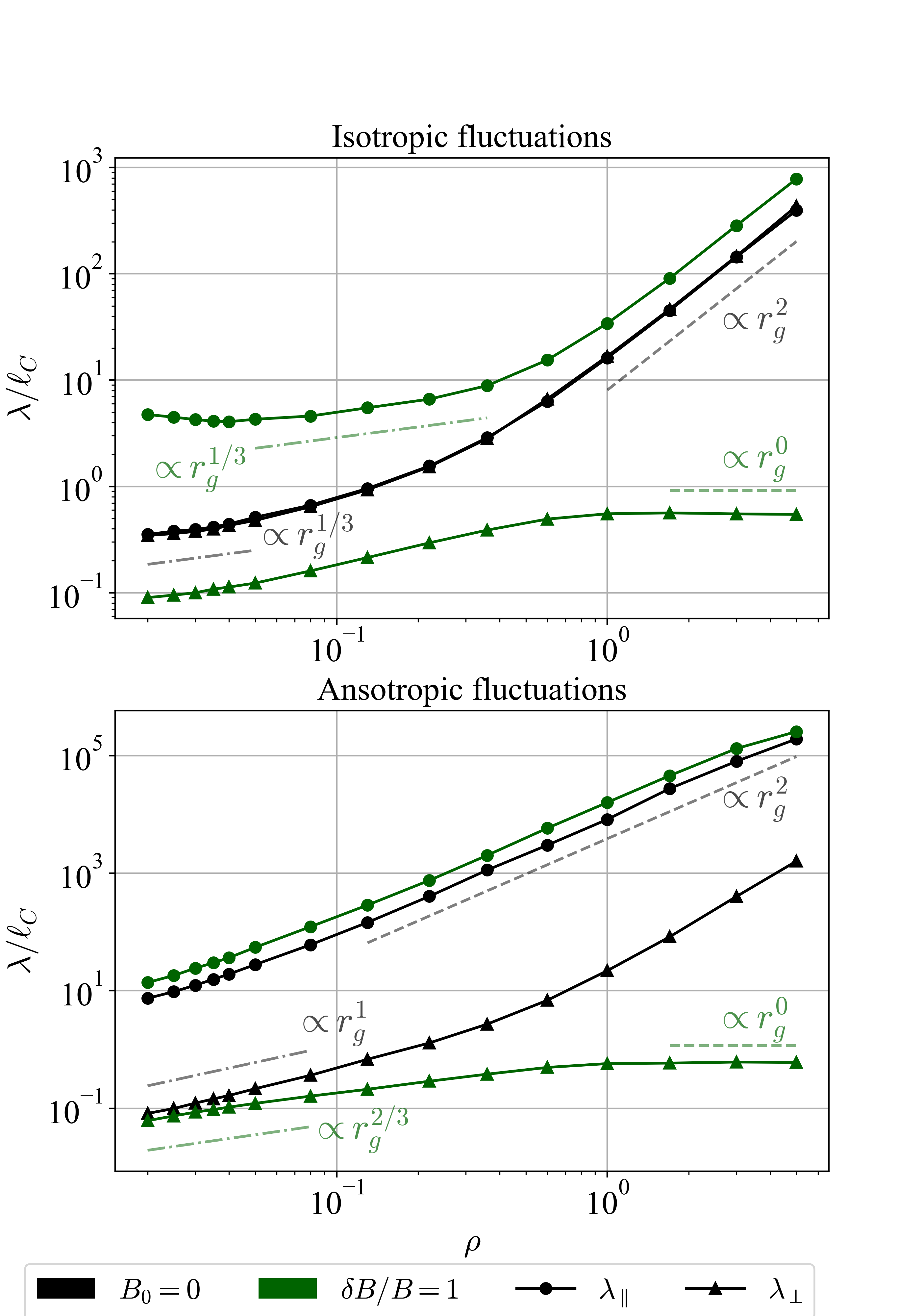}
    \caption{Parallel and perpendicular mean free path as a function of normalized rigidity for isotropic (top panel)and anisotropic (bottom panel) turbulence, with and without a uniform background field $B_0$.}
    \label{fig:Buni}
\end{figure}

Next, we focus on the effect of a background field $B_0$ on particle scattering, with and without anisotropy, with $\delta B/B_0 = 1$. The results are shown in \ffig\ref{fig:Buni}, where the black lines correspond to the same ones shown in the previous analysis (as in \ffig\ref{fig:nob0}) and are plotted here for comparison. The comparison between $\delta B / B \to \infty$ (in the sense that this corresponds to no uniform guide field $B_0=0$) and $\delta B / B = 1$ allows us to (at least partially) address the hypothesis formulated by \citet{Lemoine2023} according to which even in fields with $\delta B / B \lesssim 1$ high-energy particle transport may be influenced by small-scale magnetic-field line bends, in combination with the usual mechanism of pitch-angle diffusion present in strong-guide-field cases. The results shown in \ffig\ref{fig:Buni} show evidence of both mechanisms acting at the same time: $\lambda_{\parallel}$ in the anisotropic field shows the same scaling in both cases, suggesting that also in the finite $\delta B/B$ case, scattering is set by a typical length other than the turbulence correlation length. The distribution of $K_{\parallel}$ and $K_{\perp}$ for the $\delta B /B = 1$ case (not shown here) yields very similar results to the purely turbulent cases, in agreement with our hypothesis that the curvature scale is responsible for the scattering process. At the same time, the flattening of $\lambda_{\perp}$ at high rigidities, both in the isotropic and anisotropic case, is in better agreement with studies primarily conducted in the strong-guide-field picture \citep{nlgc2003, minnie2009}. The results shown in \ffig\ref{fig:Buni} points to the presence of an interplay between different scattering processes and support the idea that field-line curvature is an essential aspect to include in transport theories, at least for significantly turbulent fields. We leave for future work a more in-depth study focused on the careful examination of particle trajectories (e.g.\ in line with the work done by \citealt{lubke25}), as well as a direct quantification of pitch-angle scattering, which is required in order to confirm such interplay of effects and to fully understand the different mechanisms at play, the scales at which they act, and their specific role in the scattering process.

\section{Conclusion}
We performed test-particle simulations of particle transport in high-resolution synthetic turbulent fields cheaply generated with \bxc. The model's flexibility allowed us to analyze, for the first time in a synthetic cube, particle transport in fields that are intermittent, hierarchically structured, and with a controllable level of anisotropy. Anisotropy, in the context of synthetic turbulence, has been studied mainly in reference to the Goldreich--Sridhar model \citep{GS95}, or to composite models featuring a 1D slab and a 2D perpendicular component. These are all representative of highly anisotropic environments (e.g.\ the solar wind). Other anisotropy models have been studied by means of MHD simulations, whose computational cost makes a more in-depth study of this feature more difficult to achieve. Here we have focused on small anisotropy in large-amplitude magnetic fields, which have been less investigated than their strong-guide field counterparts but can be found e.g.\ in regions of the Milky Way \citep{MW2, MW1}. Our results support the need to further investigate these environments, as already suggested by \citet{kempski}, because particle transport shows characteristics that drastically differ from those of strong-guide-field environments. 

After validating our model against previous studies on isotropic fields without a background field, we showed how the introduction of anisotropy shifts the expected regimes of transport. We conjecture that such a shift is due to the fact that the turbulence correlation length is not primarily responsible for the scattering in this regime, and that field-line curvature might play a fundamental role instead. We found strong indications that when the curvature scale is taken as the reference length scale, the transport regimes found in this study are more consistent with the standard physical interpretation of the scattering process. We further extended our study to probe the validity of this interpretation also in the presence of a nonnegligible guide field ($\delta B / B = 1$), and we found signatures of an interplay of scattering mechanisms associated with strong- and weak-guide-field environments, in agreement with suggestions by \citet{Lemoine2023}. 

We showed that the \bxc{} toolkit, coupled with the MPI-AMRVAC test-particle module, provides a flexible, realistic, and cheap framework to investigate particle transport in various astrophysical settings. Specifically, the results shown in this Letter call for a deeper investigation of the different scattering mechanisms and their respective relevance at different scales and different turbulence levels, for which \bxc{} provides an affordable and reliable operational framework. Future work can study specific realistic, local conditions in order to conclusively determine the properties and scales of the scattering process in a wide variety of astrophysical environments.

\begin{acknowledgments}
We would like to thank Francesco Pucci, Nicolas Wijsen, Gene Gorbunov, Martin Lemoine, and Oreste Pezzi for useful discussions throughout the development of this work.
DM and RK acknowledge funding from the KU Leuven C1 project C16/24/010 UnderRadioSun and the Research Foundation Flanders FWO project G0B9923N Helioskill. 
FB acknowledges support from the FED-tWIN programme (profile Prf-2020-004, project ``ENERGY''), issued by BELSPO, and from the FWO Junior Research Project G020224N granted by the Research Foundation -- Flanders (FWO).
The computational resources and services used in this work were provided by the VSC (Flemish Supercomputer Center), funded by the Research Foundation Flanders (FWO) and the Flemish Government, department EWI.
\end{acknowledgments}

\begin{contribution}

All authors contributed equally.


\end{contribution}

\software{\bxc{} \citep{durrive, bxc},  MPI-AMRVAC \citep{amrvac3, amrvac4}}

\bibliography{sample701}{}
\bibliographystyle{aasjournalv7}

\end{document}